# Magnetization dynamics driven by angle-dependent spin-orbit spin transfer torque

Seo-Won Lee[1] and Kyung-Jin Lee[1,2*]

[1]*Department of Materials Science and Engineering, Korea University, Seoul 136-701, Korea*

[2]*KU-KIST Graduate School of Converging Science and Technology, Korea University, Seoul 136-701, Korea*

*Corresponding email: kj_lee@korea.ac.kr

**Spin-orbit torque allows an efficient control of magnetization by in-plane current. Recent experiments found strong angular dependence of spin-orbit torque. We theoretically investigate magnetization switching and domain wall motion in perpendicularly magnetized layers, induced by angle-dependent spin-orbit torque. We obtain analytic expressions of switching current and domain wall velocity, in agreement with numerical results. Based on the expressions, we find that the spin-orbit torque increasing with the polar angle of magnetization is beneficial for both switching and domain wall motion. Our result will serve as a guideline to design and interpret switching and domain wall experiments based on spin-orbit torque.**

In bilayers composed of ferromagnet (FM) and heavy metal (HM) with strong spin-orbit coupling, an in-plane current can drive magnetization dynamics, which is caused by spin-orbit spin transfer torque (SOT). The SOT allows an efficient manipulation of perpendicular magnetization [1] and magnetic domain wall (DW) motion [2-4] by an in-plane current.

The SOT has several advantages, compared to the conventional spin transfer torque (STT). A distinguishing advantage of SOT switching is that a write current does not pass through a tunnel barrier in magnetic tunnel junctions. This feature makes the interference among read, write, and breakdown voltages easy to manage for the applications based on magnetic random access memory (MRAM). Another important advantage is that SOT-induced perpendicular magnetization switching can be much faster than STT switching as the switching is completed within only half a precession [5, 6]. This fast switching offers a possibility of SOT-MRAM to be employed in embedded memory applications. The SOT is also attractive for DW devices because it allows very fast DW motion when combined with the Dzyaloshinskii-Moriya interaction (DMI) [7, 8]. As the interfacial DMI is always present at the FM/HM interface [9, 10], this feature offers fast operation of DW devices. Density-wise, on the other hand, SOT-based devices may not be suitable for ultrahigh density memory applications since they are in principle three-terminal devices. However, as cache memories are not of ultrahigh density, the advantages of SOT-based devices mentioned above are still very attractive.

Because of the outstanding advantages of SOT-based devices, the SOT is currently of considerable interest. For the device application, detailed understanding of SOT-induced magnetization dynamics is of critical importance. The origin of SOT has been proposed as bulk spin Hall effect in HM [11, 12] and/or interfacial spin-orbit coupling effect at FM/HM interface [13-19]. The dominant source of SOT is currently a subject under extensive investigation [20-24]. The SOT is commonly decomposed into two mutually orthogonal vector components; i.e., a field-like torque and a damping-like torque, given as

$$\boldsymbol{\tau}_{SOT} = \tau_f \hat{\mathbf{m}} \times \hat{\mathbf{y}} + \tau_d \hat{\mathbf{m}} \times (\hat{\mathbf{m}} \times \hat{\mathbf{y}}), \tag{1}$$

where $\hat{\mathbf{m}} = (\cos\varphi\sin\theta, \sin\varphi\sin\theta, \cos\theta)$ is the unit vector along the magnetization of FM, $\hat{\mathbf{y}}$ is the unit vector perpendicular to the direction of current ($\hat{\mathbf{x}}$) in the film plane, $\tau_f$ and $\tau_d$ are the magnitudes of field-like and damping-like SOTs, respectively. Recent experiments revealed an interesting feature of SOT, which is the strong dependence of $\tau_f$ and $\tau_d$ on the magnetization polar angle $\theta$ [25, 26]. We note that theories for the bulk spin Hall effect based on a drift-diffusion model or Boltzmann transport equation [23] predict no angular dependence of $\tau_f$ and $\tau_d$. On the other hand, one of us has recently reported that the interfacial spin-orbit coupling can result in such strong angular dependence of $\tau_f$ and $\tau_d$, which is caused by combined effects of the magnetization-angle dependent Fermi surface distortion and also the Fermi sea contribution [27]. Therefore, the angular dependence may shed light on the dominant origin of SOT. More importantly, this strong angular dependence of SOT should affect the performance of SOT-active devices. Therefore, the effect of angle-dependent SOT on the magnetization dynamics is an important subject to investigate. Especially, the analytic expressions of the switching current and the domain wall velocity are essential to design SOT-active devices and to interpret experimental results.

In this work, we provide such analytic expressions for the switching current and DW velocity driven by angle-dependent SOT. For the SOT-induced switching, we focus on the coherent switching mode that is valid for the cell size smaller than about 30 nm [28, 29]. We ignore $\tau_f$ for simplicity even though $\tau_f$ has a role in the perpendicular magnetization switching [29-31], because the main driving torque for the magnetization switching and DW motion is the damping-like SOT ($\tau_d$). From Ref. [25], the general form of damping-like SOT can be written as,

$$\boldsymbol{\tau}_d = \tau_{d0}\hat{\mathbf{m}} \times \left[ (\hat{\mathbf{m}} \times \hat{\mathbf{y}}) + \hat{\mathbf{z}}(\hat{\mathbf{m}} \cdot \hat{\mathbf{x}})(\chi_2 + \chi_4(\hat{\mathbf{m}} \times \hat{\mathbf{z}})^2) \right], \qquad (2)$$

where $\chi_i$ is the coefficient describing the angular dependence of $\tau_d$ (see Fig. 1). As a result, magnetization dynamics driven by damping-like SOT is described by the Landau-Lifshitz-Gilbert (LLG) equation given as

$$\frac{\partial \hat{\mathbf{m}}}{\partial t} = -\gamma \hat{\mathbf{m}} \times \mathbf{H}_{eff} + \alpha \hat{\mathbf{m}} \times \frac{\partial \hat{\mathbf{m}}}{\partial t} + \gamma \tau_{d0} \hat{\mathbf{m}} \times \left[ (\hat{\mathbf{m}} \times \hat{\mathbf{y}}) + \hat{\mathbf{z}}(\hat{\mathbf{m}} \cdot \hat{\mathbf{x}})(\chi_2 + \chi_4(\hat{\mathbf{m}} \times \hat{\mathbf{z}})^2) \right], \qquad (3)$$

where $\gamma$ is the gyromagnetic ratio, $\mathbf{H}_{eff}$ is the effective magnetic field including an in-plane field $H_x$ (i.e., required for the deterministic SOT switching [1]) and effective anisotropy field $H_{K,eff}$, $\alpha$ is the Gilbert damping parameter, $\tau_{d0}$ ($=(\hbar/2e)(\theta_{SH}J/M_S t_{FM})$) is the magnitude of damping-like SOT, $\theta_{SH}$ is the effective spin Hall angle, $M_S$ is the saturation magnetization, $t_{FM}$ is the thickness of FM, and $J$ is the current density.

We first show the analytic expression of threshold current for the SOT switching. The threshold current is obtained from a static solution of Eq. (3) with $\varphi = 0$ [5], which leads to

$$H_x \cos\theta - H_{K,eff} \cos\theta \sin\theta + \tau_{d0}(1 + \chi_2 \sin^2\theta) = 0. \qquad (4)$$

Here we assume $\chi_4 = 0$ as Eq. (3) with nonzero $\chi_4$ is not analytically solvable for magnetization switching. The threshold $\tau_{d0}$ is then determined by the condition that there is no $\theta$ satisfying Eq. (4) [5]. From this stability condition and assuming $\chi_2 \ll 1$, one finds the threshold $\tau_{d0}$ (= $\tau_{d0}^{sw}$) as,

$$\tau_{d0}^{sw} = H_{K,eff} \left( \frac{\sqrt{\Gamma - h^4 + 20h^2}}{4\sqrt{2}} + \frac{h^6 + (h^5 - 4h^3 + 12h)\sqrt{8 + h^2} + 12h^2 - 16}{16\sqrt{2}\sqrt{\Gamma - h^4 + 20h^2}} \chi_2 \right), \qquad (5)$$

where $h = H_x / H_{K,eff}$ and $\Gamma = 8 - h(8+h^2)^{3/2}$. When $h \ll 1$, Eq. (5) is further simplified as

$$\tau_{d0}^{sw} = \left(\frac{H_{K,eff}}{2} - \frac{H_x}{\sqrt{2}}\right) - \frac{1}{4}\left(H_{K,eff} - \frac{H_x}{\sqrt{2}}\right)\chi_2. \tag{6}$$

We note that the threshold value with $\chi_2 = 0$ reproduces the result of Ref. [5]. Eq. (6) shows that a positive (negative) $\chi_2$ results in the decrease (increase) of the switching current.

We next show the analytic expression of DW velocity, driven by angle-dependent damping-like SOT. For the DW motion, we consider the effective magnetic field including Dzyaloshinskii-Moriya interaction (DMI) as,

$$\mathbf{H}_{eff} = \frac{2A_{ex}}{M_S}\frac{\partial^2 \hat{\mathbf{m}}}{\partial x^2} + \frac{2K_{eff}}{M_S}m_z\hat{\mathbf{z}} + \frac{2K_d}{M_S}m_y\hat{\mathbf{y}} - \frac{2D_0}{M_S}\left(y \times \frac{\partial \hat{\mathbf{m}}}{\partial x}\right), \tag{7}$$

where $A_{ex}$ is the exchange stiffness constant, $K_{eff}$ is the effective easy-axis anisotropy, $K_d$ is the hard-axis anisotropy, and $D_0$ is the DMI constant.

By using the procedure developed by Thiele [32], we derive the equation of motion for two collective coordinates, $q(t)$ and $\varphi(t)$ from the LLG equation (Eq. (3)) in the rigid DW limit. With the Walker profile of DW [33], i.e, $m_x = \cos\varphi(t)\sin\theta$, $m_y = \sin\varphi(t)\sin\theta$, $m_z = \cos\theta$ with $\theta = 2\tan^{-1}[\exp[(x-q(t))/\lambda]]$, $q(t)$ being the DW position, $\varphi(t)$ being the DW angle tilted from $+x$-axis, and $\lambda$ being the DW width, we obtain

$$\frac{\alpha}{\lambda}\dot{q} + \dot{\varphi} = \gamma\frac{\pi}{2}\tau_{d0}\left(1 + \frac{\chi_2}{2} + \frac{3}{8}\chi_4\right)\cos\varphi,$$
$$\frac{1}{\lambda}\dot{q} + \alpha\dot{\varphi} = -\gamma\frac{\pi}{2}\frac{D_0}{\lambda M_S}\sin\varphi + \gamma\frac{K_d}{M_S}\sin 2\varphi, \tag{8}$$

where $\dot{O} = \partial O/\partial t$. With small angle approximation ($\varphi \ll 1$), the DW velocity $v_{DW}$ in the steady-state motion ($\dot{\varphi} = 0$) is given as,

$$v_{DW} = \gamma \frac{\pi}{2} \frac{\lambda D_0 \tau_{d0}\left(1+\frac{\chi_2}{2}+\frac{3}{8}\chi_4\right)}{\sqrt{D_0^2 \alpha^2 + M_S^2 \tau_{d0}^2 \left(1+\frac{\chi_2}{2}+\frac{3}{8}\chi_4\right)^2 \lambda^2}}. \tag{9}$$

Eq. (9) shows that the angular dependence with positive $\chi_2$ and $\chi_4$ increases the DW velocity. Eq. (5) for the SOT switching and Eq. (9) for the SOT-induced DW motion are our central results in this work.

To verify the validity of these analytic expressions, we numerically compute the threshold switching current and DW velocity for various angular dependences of damping-like SOT. First, we compute the switching current ($I_{SW}$) as a function of $\chi_2$ and $H_x$ based on macrospin simulations (Fig. 2). We find that the analytic expression for the switching current (Eq. (5)) is in reasonable agreement with numerical results. Some deviations between Eq. (5) and numerical results are observed for a large $|\chi_2|$, which is caused by the assumption of $\chi_2 \ll 1$ in the analytic derivation.

We next compare the analytic expression of DW velocity (Eq. (10)) and numerical results (Fig. 3). We find that Eq. (9) is in reasonable agreement with numerical results for a small current. The deviation between Eq. (9) and numerical results becomes large for a large current (Fig. 3(b)). This deviation has two sources. One is that the DW angle $\varphi$ becomes large with increasing current (Fig. 3(c)) so that the small angle approximation used for the derivation of Eq. (9) becomes invalid. The other is that the SOT also tilts the domain parts (Fig. 3(d)). For the up domain (// $z$-axis), the effective field caused by damping-like SOT is in the direction of $\hat{\mathbf{m}} \times \hat{\mathbf{y}} (= \hat{\mathbf{z}} \times \hat{\mathbf{y}} \equiv \hat{\mathbf{x}})$ so that the magnetization tilts along the +x direction. The opposite domain on the other side of DW tilts along the –x direction. This domain tilting becomes larger for a larger current, so that the Walker profile of DW becomes invalid [34]. We note however that Eq. (9) describes the overall behavior of DW velocity reasonably well in wide ranges of current because the deviation is not very substantial (Fig. 3(b)). We also note that there could be DW surface tilting due to DMI in realistic two dimensional systems [35] although it is not captured by our one dimensional model or simulation. However, its effect on the DW velocity is negligible in the small angle approximation used for the derivation of Eq. (9).

Finally, we discuss the implications of our result. We show that the angular dependence of SOT with a positive $\chi_i$ provides a lower switching current and faster DW motion, which are also verified by numerical results. According to our previous work on the angular dependence of SOT [27], a larger ratio of $\alpha_R k_F$ to $J_{sd}$ causes not only a larger overall magnitude of damping-like SOT but also a larger $\chi_i$, where $\alpha_R$ is the Rashba constant describing the magnitude of interfacial spin-orbit coupling, $k_F$ is the Fermi wavevector, and $J_{sd}$ is the sd exchange coupling. Therefore, our result suggests that material engineering to enhance $\alpha_R k_F/J_{sd}$ will be beneficial for the application of SOT-MRAM and SOT-based DW devices. Furthermore, our result indicates that the angular dependence of SOT should be taken into account to estimate the effective spin Hall angle from the switching or DW experiments. For this purpose, one can apply our analytic expressions to interpret the switching or DW experiments with measuring the SOT in wide ranges of the magnetization angle as done in Refs. [25, 26].

To summarize, we investigate the effect of angular dependence of damping-like SOT on perpendicular magnetization switching and DW motion. We derive explicit analytic expressions for switching current and DW velocity, which are in good agreement with numerical results. With these analytic expressions, we find that the damping-like SOT, which increases with the polar angle $\theta$ of magnetization, reduces the switching current and enhances the DW velocity. Our result will be of importance for the applications since it can be used to estimate essential physical quantities such as the effective spin Hall angle and to design practical SOT-active devices.

This work was supported by the National Research Foundation of Korea (NRF) (NRF-2013R1A2A2A01013188, 2011-0028163).

**REFERENCE**

1. I. M. Miron *et al*., Nature **476**, 189 (2011).

2. A. Thiaville, S. Rohart, É. Jué, V. Cros, and A. Fert, Europhys. Lett. **100**, 57002 (2012).

3. S. Emori *et al*., Nature Mat. **12**, 611 (2013).

4. K.-S. Ryu *et al*., Nature Nanotechnol. **8**, 527 (2013).

5. K.-S. Lee, S.-W. Lee, B.-C. Min, and K.-J. Lee, Appl. Phys. Lett. **102**, 112410 (2013).

6. K.-S. Lee, S.-W. Lee, B.-C. Min, and K.-J. Lee, Appl. Phys. Lett. **104**, 072413 (2014).

7. I. E. Dzialoshinskii, Sov. Phys. JETP **5**, 1259 (1957).

8. T. Moriya, Phys. Rev. **120**, 91 (1960).

9. A. Fert and P. M. Levy, Phys. Rev. Lett. **44**, 1538 (1980).

10. K.-W. Kim, H.-W. Lee, K.-J. Lee, and M. D. Stiles, Phys. Rev. Lett. **111**, 216601 (2013).

11. Y. A. Bychkov and E. I. Rashba, JETP Lett. **39**, 78 (1984).

12. J. E. Hirsch, Phys. Rev. Lett. **83**, 1834 (1999).

13. A. Manchon and S. Zhang, Phys. Rev. B **78**, 212405 (2008).

14. K. Obata and G. Tatara, Phys. Rev. B **77**, 214429 (2008).

15. X. Wang and A. Manchon, Phys. Rev. Lett. **108**, 117201 (2012).

16. K.-W. Kim, S.M. Seo, J. Ryu, K.-J. Lee, and H.-W. Lee, Phys. Rev. B **85**, 180404 (2012).

17. D. A. Pesin and A. H. MacDonald, Phys. Rev. B **86**, 014416 (2012).

18. P. M. Haney, H.-W. Lee, K.-J. Lee, A. Manchon, and M. D. Stiles, Phys. Rev. B **88**, 214417 (2013).

19. H. Kurebayashi *et al*., Nature Nanotechnol. **9**, 211 (2014).

20. L. Liu *et al*., Science **336**, 555 (2012).

21. L. Liu, O. J. Lee, T. J. Gudmundsen, D. C. Ralph, and R. A. Buhrman, Phys. Rev. Lett. **109**, 096602 (2012).

22. J. Kim *et al.*, Nature Mat. **12**, 240 (2013).

23. P. M. Haney, H. W. Lee, K. J. Lee, A. Manchon, and M. D. Stiles, Phys. Rev. B **87**, 174411 (2013).

24. X. Fan *et al.*, Nat. Commun. 5:3042, doi: 10.1038/ncomms4042 (2014).

25. K. Garello *et al.*, Nature Nanotechnol. **8**, 587 (2013).

26. X. Qiu *et al.*, Sci. Rep. 4, 4491; DOI:10.1038/srep04491 (2014).

27. K.-S. Lee *et al.*, Phys. Rev. B **91**, 144401 (2015).

28. K. Garello *et al.*, Appl. Phys. Lett. **105**, 212402 (2014).

29. C. Zhang *et al.*, Appl. Phys. Lett. **107**, 012401 (2015).

30. T. Taniguchi, S. Mitani, and M. Hayashi, Phys. Rev. B **92**, 024428 (2015).

31. W. Legrand, R. Ramaswamy, R. Mishra, and H. Yang, Phys. Rev. Appl. **3**, 064012 (2015).

32. A. A. Thiele, Phys. Rev. Lett. **30**, 230 (1973).

33. N. L. Schryer and L. R. Walker, J. Appl. Phys. **45**, 5406 (1974).

34. S. Emori *et al.*, Phys. Rev. B **90**, 184427 (2014).

35. O. Boulle *et al.*, Phys. Rev. Lett. **111**, 217203 (2013).

**Figure legends**

Figure 1. Angular dependence of damping-like SOT depending on $\chi_2$ and $\chi_4$.

Figure 2. Switching current $I_{SW}$ as a function of (a) $\chi_2$ and (b) $H_x$. Symbols are numerical results whereas lines are the analytic result (Eq. (5)). Following parameters are used: the diameter of circular shaped free layer = the width of HM layer = 30 nm, the thickness of HM layer = 2 nm, $t_{FM}$ = 1 nm, $\alpha$ = 0.01, $M_S$ = 1000 emu/cm$^3$, and $\theta_{SH}$ = 0.3.

Figure 3. Domain wall velocity as a function of (a) $\chi_2$ and (b) in-plane current. In (a) and (b), symbols are numerical results whereas lines are the analytic result (Eq. (9)). (c) Domain wall angle $\varphi$ tilted from *x*-axis with increasing current and (d) *x*-component of magnetization along *x*-axis at a low current ($I$ = 6 µA corresponding to current density of 6 × 10$^6$ A/cm$^2$) and a high current (30 µA). Following parameters are used: the width of HM layer = 50 nm, the thickness of HM layer = 2 nm, $t_{FM}$ = 1 nm, perpendicular anisotropy energy density $K_u$ = 1.2 × 10$^7$ erg/cm$^3$, $\alpha$ = 0. 1, $M_S$ = 1000 emu/cm$^3$, $\theta_{SH}$ = 0.3, $A_{ex}$ = 1.6 × 10$^{-6}$ erg/cm, and $D_0$ = 0.5 erg/cm$^2$.

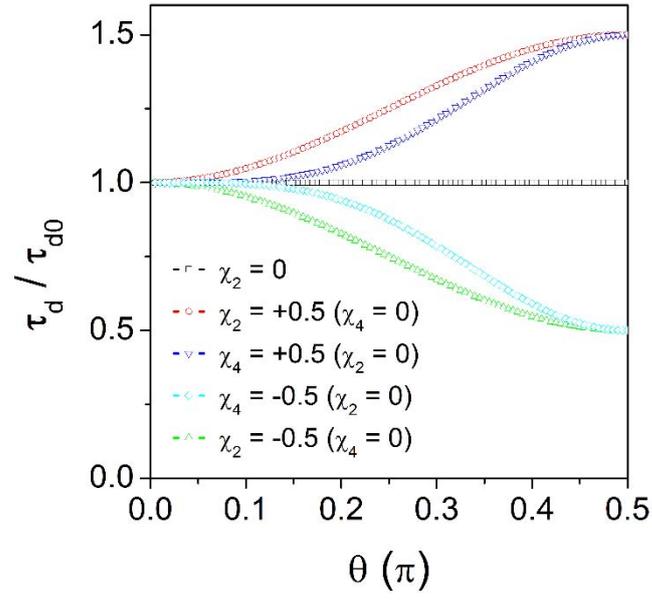

Figure 1. Angular dependence of damping-like SOT depending on $\chi_2$ and $\chi_4$.

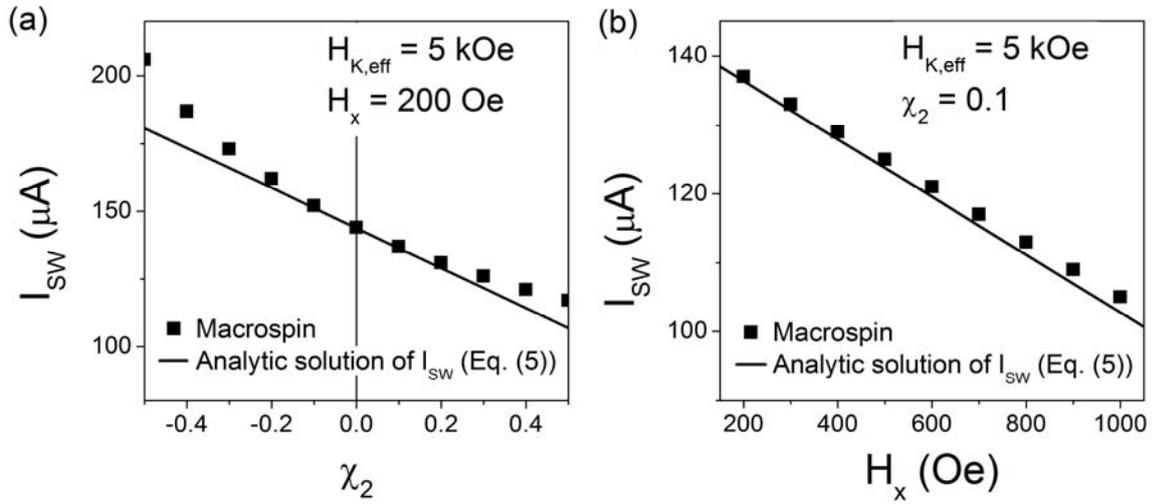

Figure 2. Switching current I$_{SW}$ as a function of (a) $\chi_2$ and (b) $H_x$. Symbols are numerical results whereas lines are the analytic result (Eq. (5)). Following parameters are used: the diameter of circular shaped free layer = the width of HM layer = 30 nm, the thickness of HM layer = 2 nm, $t_{FM}$ = 1 nm, $\alpha$ = 0.01, $M_S$ = 1000 emu/cm³, and $\theta_{SH}$ = 0.3.

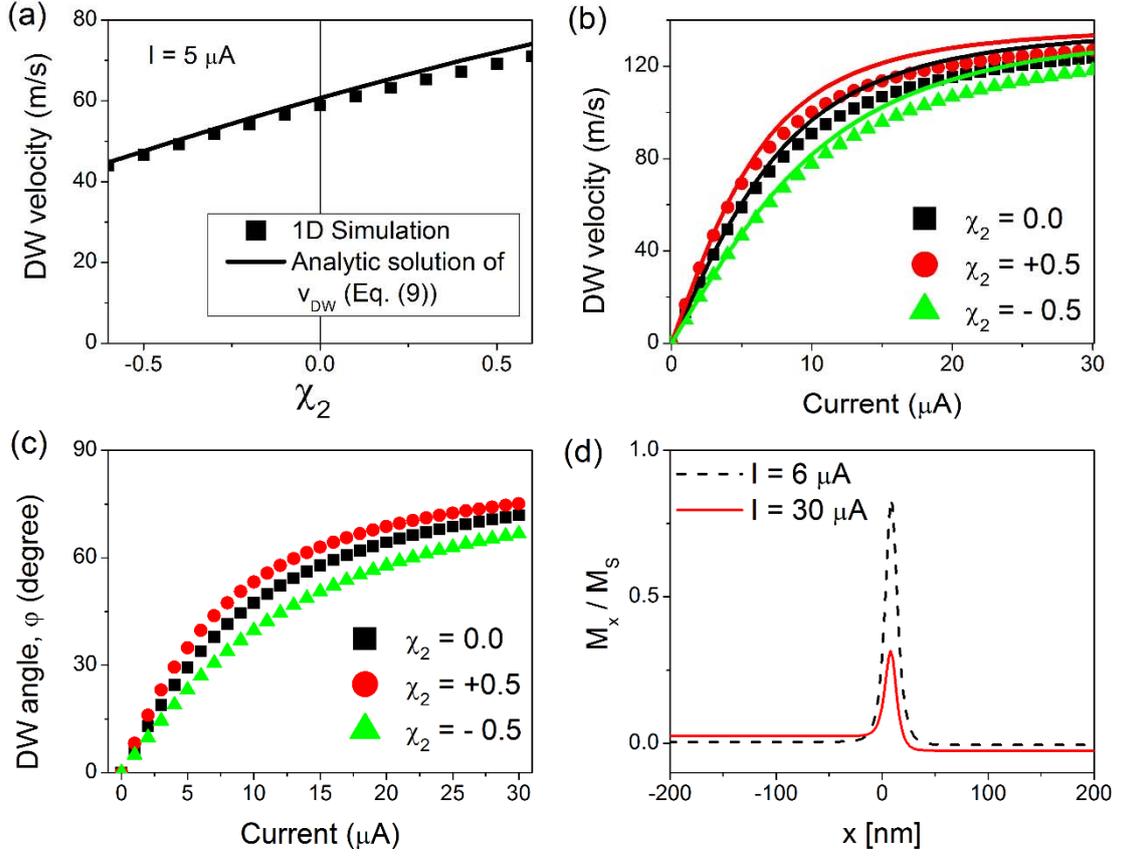

Figure 3. Domain wall velocity as a function of (a) $\chi_2$ and (b) in-plane current. In (a) and (b), symbols are numerical results whereas lines are the analytic result (Eq. (9)). (c) Domain wall angle $\varphi$ tilted from $x$-axis with increasing current and (d) $x$-component of magnetization along $x$-axis at a low current ($I = 6$ μA corresponding to current density of $6 \times 10^6$ A/cm$^2$) and a high current (30 μA). Following parameters are used: the width of HM layer = 50 nm, the thickness of HM layer = 2 nm, $t_{FM}$ = 1 nm, perpendicular anisotropy energy density $K_u = 1.2 \times 10^7$ erg/cm$^3$, $\alpha$ = 0.1, $M_S$ = 1000 emu/cm$^3$, $\theta_{SH}$ = 0.3, $A_{ex} = 1.6 \times 10^{-6}$ erg/cm, and $D_0$ = 0.5 erg/cm$^2$.